\documentclass[runningheads]{llncs}
\usepackage[T1]{fontenc}
\usepackage[font=small]{caption}

\usepackage{graphicx}
\usepackage{hyperref}
\usepackage{orcidlink}

\usepackage{latexsym}

\newcommand{\para}[1]{\paragraph{\textnormal{\textbf{#1}.}}}

\usepackage[titlenumbered,ruled,vlined]{algorithm2e}
\usepackage[normalem]{ulem}

\usepackage{tabularx}
\usepackage[export]{adjustbox}
\usepackage{subcaption}
\usepackage{booktabs}
\usepackage{multirow}
\usepackage{xcolor,colortbl}
\usepackage{float}
\usepackage{siunitx}

\usepackage{calrsfs}
\DeclareMathAlphabet{\pazocal}{OMS}{zplm}{m}{n}
\DeclareMathAlphabet{\pazobfcal}{OMS}{cmsy}{b}{n}

\usepackage{amssymb}
\usepackage{amsmath}

\usepackage{enumitem} 

\setitemize{noitemsep}
\setlist{topsep=0pt, leftmargin=*}

\newcommand{\uls}{\begin{itemize}[leftmargin=*]}
\newcommand{\ule}{\end{itemize}}
\newcommand{\ols}{\begin{enumerate}[leftmargin=*]}
\newcommand{\ole}{\end{enumerate}}
\newcommand{\li}{\item}

\newboolean{showcomments}
\setboolean{showcomments}{true} 
\ifthenelse{\boolean{showcomments}}
{
	\newcommand{\nb}[3]{
		{\colorbox{#2}{\bfseries\sffamily\scriptsize\textcolor{white}{#1}}}
		{\textcolor{#2}{$\blacktriangleright$\textsf\small{#3}$\blacktriangleleft$}}}
	 
}{
	\newcommand{\nb}[3]{}
} 

\begin{document}
\title{LURE-RAG: Lightweight Utility-driven Reranking for Efficient RAG}
%

\author{Manish Chandra\inst{1}\orcidlink{0009-0000-6156-5337} \and
Debasis Ganguly\inst{1}\orcidlink{0000-0003-0050-7138} \and
Iadh Ounis\inst{1}\orcidlink{0000-0003-4701-3223}}
\authorrunning{M. Chandra et al.}
%
\institute{University of Glasgow, Glasgow, United Kingdom
\email{m.chandra.1@research.gla.ac.uk, Debasis.Ganguly@glasgow.ac.uk, iadh.ounis@glasgow.ac.uk}\\
}

\maketitle              
\begin{abstract}
%
Most conventional Retrieval-Augmented Generation (RAG) pipelines rely on relevance-based retrieval, which often misaligns with utility -- that is, whether the retrieved passages actually improve the quality of the generated text specific to a downstream task such as question answering or query-based summarization. The limitations of existing utility-driven retrieval approaches for RAG are that, firstly, they are resource-intensive typically requiring query encoding, and that secondly, they do not involve listwise ranking loss during training. The latter limitation is particularly critical, as the relative order between documents directly affects generation in RAG. To address this gap, we propose Lightweight Utility-driven Reranking for Efficient RAG (LURE-RAG), a framework that augments any black-box retriever with an efficient LambdaMART-based reranker. Unlike prior methods, LURE-RAG trains the reranker with a listwise ranking loss guided by LLM utility, thereby directly optimizing the ordering of retrieved documents. Experiments on two standard datasets demonstrate that LURE-RAG achieves competitive performance, reaching 97–98\% of the state-of-the-art dense neural baseline, while remaining efficient in both training and inference. Moreover, its dense variant, UR-RAG, significantly outperforms the best existing baseline by up to 3\%.

\keywords{RAG, listwise ranking, lightweight reranker}
\end{abstract}
\section{Introduction}

Large Language Models (LLMs) have shown strong ability to generate fluent and often factually grounded text, yet they are limited by their parametric knowledge (which is fixed at training time) \cite{rag,rag_techniques}. They may also hallucinate or fail on domain‐specific or newly emerging information. Retrieval-Augmented Generation (RAG) has emerged as a prominent framework to address these limitations by combining external retrieval of relevant documents with language model generation \cite{replug,ragReview}. RAG systems first retrieve documents or passages from some corpus given an input, then augment the LLM’s input with those documents, and finally generate a response grounded on both retrieved and internal knowledge. RAG has become a core paradigm in knowledge‐intensive NLP tasks \cite{huangsurveyRAG,guptasurveyRAG}. RAG systems usually consist of multiple modules including at least a retriever and a generator. Some systems may have other modules to further enhance effectiveness on downstream tasks, like a reranker \cite{re2g} or a decision maker deciding when to retrieve \cite{adaptiveRAG,smartrag}.


In many RAG pipelines \cite{rag_techniques,ragReview}, relevance is defined in the traditional IR sense, such as lexical or semantic similarity between documents and the query. However, in the RAG setting, semantic relevance to the query alone does not necessarily translate into utility for the LLM’s downstream generation~\cite{relevance_propagated,power_of_noise,llm_utility_judgement}, that is, whether the retrieved documents actually help the model produce more accurate, coherent, or useful answers. This discrepancy arises because, even when documents are individually relevant, concatenating them into a single context may introduce incoherence or inconsistency, which can in turn mislead the LLM and degrade the quality of the generated output.
%
%
Moreover, even if a document is semantically similar to an input question, it may miss crucial facts required for the correct answer generation.
Recent works \cite{smartrag,replug} showed that using utility-driven signals (e.g. metrics derived from LLM outputs and the ground-truth answers) as supervision can better align retrieval (and subsequent reranking in some cases) to what improves generation.

Despite the progress in the direction of utility-driven signals, there remain important gaps. Firstly, the loss functions used in the retriever or reranker training do not reflect order or comparative ranking (i.e., a document in rank 1 vs rank 3 matters). This formulation treats all mis-rankings equally, failing to reflect the importance of ranking errors at the top of the list versus those deeper in the ranking. It has indeed been shown that the order of documents matter in RAG \cite{liu-etal-2024-lost,power_of_noise}. Secondly, the existing utility‐driven methods are not efficient in terms of training and inference cost. To the best of our knowledge, there is no existing work focusing on lightweight, efficient reranking methods that still integrate utility supervision in a RAG setup.

In this work, we propose a framework that treats the retriever and the LLM as black-boxes, and adds a lightweight reranker on top of the retrieved candidates, which is efficient both for training and inference. We train the reranker using listwise ranking loss derived from supervision given by the LLM utility, i.e., how much each document/passage contributes to the quality of generation. We call our proposed framework \textbf{LURE-RAG} (Lightweight Utility-driven Reranking for Efficient RAG).

\para{Our Contributions}
Our main contributions are summarized below.
\begin{itemize}
    \item We propose a \textit{black‐box retriever, lightweight reranker and black-box generator} architecture for RAG, called LURE-RAG.
    \item We propose a listwise ranking‐loss based training scheme for the reranker that uses LLM utility signals (answer correctness) to supervise which retrieved documents should be ranked higher.
    \item We demonstrate that LURE-RAG shows a competitive performance, coming within 97-98\% of the strongest dense neural reranker baseline, while being computationally efficient with modest training and inference overhead, making it practical for real‐world RAG systems.
    \item We demonstrate that the dense neural counterpart of LURE-RAG, which we call UR-RAG (Utility-driven Reranking for RAG), significantly outperforms competing baselines on multiple knowledge‐intensive tasks.
\end{itemize}

\section{Related Work}
In this section, we position LURE-RAG within the literature on RAG, distinguishing between those methods that do not explicitly use utility-driven supervision, and those that do. We then cover learning to rank using listwise or pairwise ranking losses, since LURE-RAG leverages those approaches.

\subsection{Retrieval Augmented Generation}
RAG \cite{rag} refers broadly to systems which combine a retrieval component (obtaining a ranked list of the most likely relevant documents from a collection) together with a generative (LLM) component that uses those retrieved passages to output answers. RAG helps address LLM limitations like outdated knowledge, hallucinations, or inability to access domain-specific or emerging information \cite{rag,ragReview}.

\para{Relevance-based RAG}
Early approaches to RAG rely primarily on the traditional IR notion of relevance, rather than explicitly optimizing for downstream LLM utility \cite{rag_techniques,ragReview}. In these methods, retrieval is guided by query-document lexical or semantic similarity, often using sparse methods such as BM25 \cite{bm25} or dense dual-encoder retrievers \cite{dpr}. Once retrieved, the top-$k$ passages are either directly provided to the generator \cite{rag,aicl}, or fused using models such as Fusion-in-Decoder (FiD) \cite{izacard2021fid}.

\para{Supervised Retriever (Utility-driven RAG)}
These methods incorporate a downstream LLM output utility (e.g., the correctness of answers in a factoid QA task) into the training of retriever or reranker. A noticeable example is RePlug \cite{replug}, which integrates a dense retriever (e.g., Contriever \cite{contriever}) into a black-box LLM pipeline. Unlike classical supervised retrievers trained on relevance labels, RePlug leverages LLM utility signals, i.e., how much a retrieved passage improves the LLM’s likelihood of producing the correct answer. This methodology aligns the retriever with the needs of the generator rather than with human-annotated relevance judgments. The training objective in RePlug relies on KL divergence between the retriever’s predicted distribution over passages and the LLM-utility-derived distribution.

While RePlug is described as a plug-in approach, in practice, it depends on dense retrievers such as Contriever. This makes it incompatible with lightweight or non-parametric retrievers such as BM25, because BM25 does not produce differentiable representations that can be trained against utility-derived soft labels. As a result, RePlug cannot directly optimize BM25 or other non-parametric retrievers, limiting its generality in resource-constrained or latency-sensitive settings. Furthermore, the KL divergence objective in RePlug does not take into account the utility-driven ranking of the documents, i.e., it treats misclassifications at the top of the list the same as those at the bottom. In contrast, our proposed method LURE-RAG makes use of a ranking loss and is compatible with any retriever since it assumes the retriever to be a black-box.

Another recent work \cite{llm_utility_judgement} highlights the misalignment between topical relevance and actual utility in RAG, proposing that LLMs can serve as zero-shot utility judges through a k-sampling listwise approach to mitigate inherent position biases. While effective, this method is computationally expensive at inference time, as it requires multiple LLM calls to shuffle and aggregate document scores. In contrast, our proposed LURE-RAG framework distills this utility-driven intelligence into a lightweight LambdaMART-based reranker, providing an efficient and scalable way to integrate utility signals without requiring multiple LLM calls. We do not adopt this approach as a baseline in our experiments since its reliance on iterative LLM calls makes it computationally prohibitive at scale, and its design (direct utility judgment rather than reranking) targets a different problem formulation than ours.


\subsection{Lightweight Learning to Rank}
Learning to Rank (LTR) in IR focuses on training models to order documents by relevance. Among the most influential approaches are RankNet \cite{ranknet}, LambdaRank \cite{lambdarank}, and LambdaMART \cite{lambdamart}.

RankNet introduced a neural network–based pairwise ranking model that learns by minimizing a cross-entropy loss between predicted pairwise preferences and ground-truth labels. RankNet demonstrated the effectiveness of using gradient-based optimization for ranking, a key contribution for subsequent neural ranking models such as \cite{pasumarthi2019tf,ai2018learning}. LambdaRank improved upon RankNet by introducing the concept of lambdas: gradient signals that depend not only on the pairwise preference but also on the impact of swapping two documents on an evaluation metric such as NDCG (Normalized Discounted Cumulative Gain). This allowed models to optimize ranking metrics more directly without needing their gradients explicitly.

LambdaMART combined LambdaRank with MART (Multiple Additive Regression Trees) \cite{mart}, a boosted tree ensemble method. By applying the lambda-based gradients to tree-based learners, LambdaMART achieved state-of-the-art results in large-scale learning-to-rank benchmarks and became widely deployed in industrial search \cite{lambdamart,Wu2010Adapting} and recommendation systems \cite{Hu2019Unbiased}. LambdaMART is particularly appealing due to its efficiency, scalability, and interpretability, which make it suitable as a lightweight reranker in real-world pipelines. In our work, we build on these strengths by employing LambdaMART within LURE-RAG as the core reranker, using LLM-derived utility signals to provide supervision. This allows us to capture task-specific document preferences while retaining the efficiency and transparency of tree-based models.

\section{Proposed Methodology} \label{sec:methodology}
\begin{figure*}[t]
\centering
\includegraphics[width=0.95\textwidth]{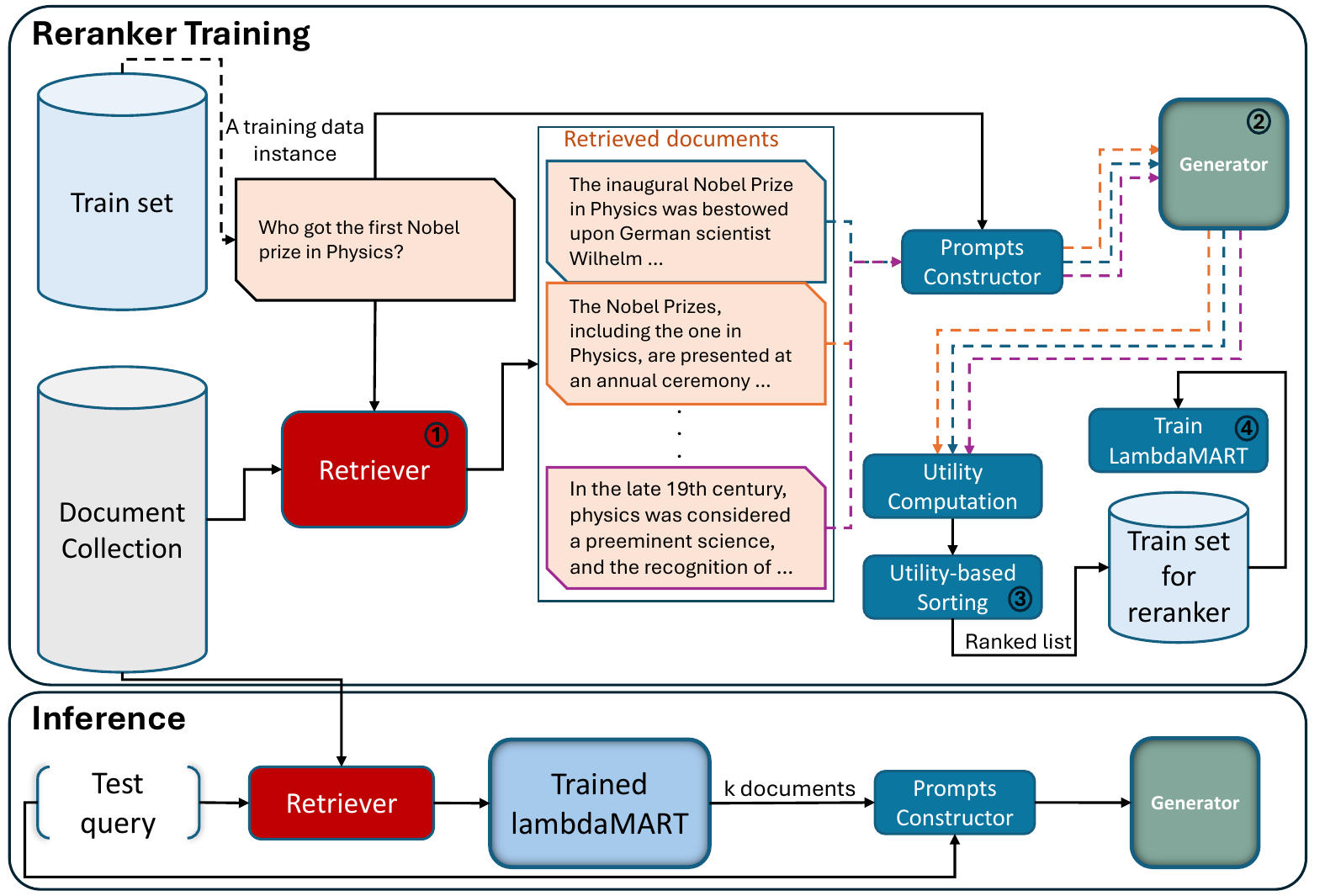}
\caption{
Schematic diagram of LURE-RAG workflow. \textcircled{1} For the given query, documents are retrieved using a (black-box) retriever. \textcircled{2} Each of these documents is used as a context for the query and the prompts are constructed. These prompts are fed into a generator one by one to get the LLM predictions. \textcircled{3} LLM's posteriors are used to compute the utility of each document. The documents are then sorted based on the utility scores and only top-$k$ of them are retained. These form the training instances for training a reranker. \textcircled{4} LambdaMART is trained using thus obtained training data.
\label{fig:arch}
}
\end{figure*}

In this section, we first provide the problem definition and then the details of our proposed strategy for lightweight utility-driven reranking for efficient RAG (LURE-RAG). Figure~\ref{fig:arch} presents an overarching view of the proposed approach.

\subsection{Problem Definition}
Let $\pazocal{Q}$ denote a set of queries (or user questions), where each query $q \in \pazocal{Q}$ is represented as a sequence of tokens $q=(w_1,\ldots,w_{|q|})$. For a given query $q$, a retriever $\pazocal{R}$ returns a candidate set of $N$ documents:
$D_q = \{d_1,\ldots,d_{N}\}, d_i \in \pazocal{D}$,
where $\pazocal{D}$ is the document collection. In a traditional retrieval setting, documents are ranked according to relevance scores. However, in RAG, what ultimately matters is the \textit{utility} of a document, i.e., how much it improves the LLM's ability to generate correct responses.

We formalize the utility-driven reranking problem as follows: given a query $q$, a set of candidate documents $D_q$, and a utility function $U(q,d)$ estimated via LLM outputs, learn a ranking function:
$f_\theta : (q,d) \mapsto \mathbb{R}$,  
such that the induced ranking maximizes the expected downstream utility. The reranker $f_\theta$ assigns a new relevance score $s_{q,d} = f_\theta(q,d)$ to each document $d \in D_q$. 

\subsection{Utility-driven Supervision}
Following prior work (\cite{replug}), we define document utility by measuring the effect of including $d$ in the LLM’s context on the downstream task-specific score (e.g., F1 or Exact Match). Formally, if $y_q$ is the gold answer and $\hat{y}_q(d)$ is the model’s output when conditioned on $d$, then
$U(q,d) = \pazocal{P}(y_q, \hat{y}_q(d))$,
where $\pazocal{P}(\cdot)$ is a task-specific evaluation metric. The utilities $U(q,d)$ serve as supervision signals for training the reranker.

\subsection{Feature Construction}
Each query–document pair $(q,d)$ is represented by a feature vector $\phi(q,d) \in \mathbb{R}^e$ such that:
$\phi(q,d) = (f_1(q,d), f_2(q,d),\ldots,f_e(q,d))$,
where $e$ is the total number of features. Specifically, we use lexical and IR-based features as follows.

\para{Query statistics}
Since longer queries may provide more discriminative context, we use the total number of query terms and the distinct number of query terms as features, i.e., $f_1 = |q|$ and $f_2 = |\{w \in q\}|$. We also use features that capture how ``rare'' or ``specific'' the query terms are, i.e., $f_3 = \min_{w \in q} IDF(w)$, $f_4 = \max_{w \in q} IDF(w)$ and $f_5 = \frac{1}{|q|}\sum_{w \in q} IDF(w)$.

\para{Document statistics}
We also use the corresponding features for documents, i.e., $f_6 = |d|$ and $f_7 = |\{w \in d\}|$. We use these features because longer documents might get unfairly favored because of their higher term frequencies. Similarly, we use the informativeness-related feature on the document side, i.e., $f_8 = \min_{w \in d} IDF(w)$, $f_9 = \max_{w \in d} IDF(w)$ and $f_{10} = \frac{1}{|d|}\sum_{w \in d} IDF(w)$.

\para{Query-Document features}
As a measure of lexical match, we use the number of overlapping terms as a feature, i.e., $f_{11} = q \cap d$. Furthermore, as a typically strong baseline for query-document matching, we use the retrieval score \cite{Liu2009LearningToRank}, i.e., if BM25 is used for retrieval, $f_{12} = BM25(q,d)$.

\para{Topic features}
Latent Dirichlet Allocation (LDA) \cite{lda} provides a lower-dimensional topical representation that captures co-occurrence patterns beyond surface word overlap. These topic-level signals help identify cases where a query and a candidate document are \textit{semantically} aligned even if they share few exact terms, making them a complementary feature to lexical statistics. The $\phi$ vector for a specific topic is the probability distribution over all words in the vocabulary and the $\theta$ vector for a specific document is the probability distribution over all topics.

Let $\theta_q \in \mathbb{R}^M$ and $\theta_d \in \mathbb{R}^M$ be the topic distributions for query $q$ and document $d$ over $M$ topics. We use cosine similarity to capture the semantic similarity beyond words overlap, i.e., $f_{13}=\theta_q\cdot\theta_d/|\theta_q||\theta_d|$. Furthermore, in order to capture the intuition that if the query's dominant topics are also well covered in the document, then the document is likely useful, we use the sum of document topic probabilities corresponding to the query’s top topics as a feature. Mathematically, $f_{14} = \sum_{m \in \text{Top}_a(\theta_q)} \theta_d(m)$, where $\text{Top}_a(\theta_q)$ denotes the top $a$ topics in the query and $\theta_d(m)$ denotes topic $m$'s probability in the document $d$.

\subsection{Listwise Learning to Rank with LambdaMART}
LambdaMART \cite{lambdamart} is a listwise learning-to-rank algorithm that combines LambdaRank’s \cite{lambdarank} gradient formulation with gradient-boosted regression trees (MART)~\cite{mart}. For each query $q$, we have a candidate document set $D_q$ with corresponding utility labels $\{U(q,d) | d \in D_q\}$. LambdaMART trains an ensemble of regression trees to predict scores $s_{q,d}$ such that the induced predicted ranking aligns with the ground-truth ranking based on the utilities. The objective is to minimize a loss that directly correlates with ranking quality. In LambdaMART, if $d_i$ is ranked higher than $d_j$ for the query $q$, the gradients are defined as follows:
\begin{equation}
    \lambda_{ij} = -\sigma \cdot \frac{1}{1 + \exp^{\sigma (s_{q,d_i} - s_{q,d_j})}} \cdot \Delta NDCG_{ij},
\end{equation}
where, $s_{q,d_i}$ and $s_{q,d_j}$ are predicted scores for documents $d_i$ and $d_j$ respectively, $\sigma > 0$ is a scaling parameter and $\Delta NDCG_{ij}$ is the change in NDCG if $d_i$ and $d_j$ were swapped. The reranker updates its parameters using these $\lambda$ values, which approximate the gradient of NDCG with respect to the predicted scores.

\subsection{Inference}
At test time, given a query, the first-stage retriever first retrieves $N$ documents/passages. The trained lambdaMART model is then used to rank them and select top $k$. These $k$ documents are then used to construct the prompt for the LLM. We note that while at training time we score each training example independently, at test time the LLM observes a prompt, i.e., a sequence of examples. We leave modeling the dependence between different training examples to future work.

\section{Experiment Setup} \label{sec:expsetup}

\subsection{Research Questions and Datasets}
We investigate the following research questions:
\uls
\li 
\textbf{RQ-1}: Does utility-driven reranking lead to a better downstream performance than traditional relevance-based retrieval across different LLMs and datasets?
\li 
\textbf{RQ-2}: Does training a reranker with listwise ranking loss derived from LLM utility lead to a better downstream performance?
\li
\textbf{RQ-3}: Can a lightweight LambdaMART-based reranker (LURE-RAG) achieve competitive performance compared to dense neural rerankers (RePlug), while being more computationally efficient?
\li
\textbf{RQ-4}: Is the reranker transferrable, i.e., can the reranker trained on one LLM's utilities perform well when used with other downstream LLMs of different sizes?

\ule

\para{Datasets}
We conduct our experiments on two standard open-domain question answering (QA) datasets, namely Natural Questions-Open (NQ-Open) and TriviaQA (see Table~\ref{tab:datasets}). More details on each dataset are as follows.

\noindent \textbf{NQ-Open (NQ)}:
The NQ-open dataset \cite{nq_open}, a subset of the NQ dataset \cite{nq}, differs by removing the restriction of linking answers to specific Wikipedia passages, thereby mimicking a
more general information retrieval scenario similar to web searches. Following the methodology of \cite{nq_open}, our primary source for answering queries is the English Wikipedia dump as of 20 December 2018. Consistent with the Dense Passage Retrieval (DPR) approach \cite{dpr}, each Wikipedia article in this dump is segmented into non-overlapping passages of 100 words.

\begin{table}[t]
\centering
\caption{Statistics of the datasets used in our experiments.}
\label{tab:datasets}
\begin{adjustbox}{width=0.62\columnwidth}  
\begin{tabular}{lcccc}
\toprule
\textbf{Dataset} & \textbf{Domain} & \textbf{\#Train} & \textbf{\#Dev} & \textbf{\#Test} \\
\midrule
NQ-Open & Google queries & 72,209 & 8,757 & 2,889 \\
TriviaQA & Trivia questions & 78,785 & 8,837 & 11,313 \\
\bottomrule
\end{tabular}
\end{adjustbox}
\end{table}

\noindent \textbf{TriviaQA (TQA)}:
The TriviaQA dataset \cite{triviaqa} is a large-scale benchmark consisting of question–answer pairs originally collected from trivia enthusiasts and quiz-league websites. Each question is accompanied by one or more evidence documents, either from Wikipedia or from web search results. For open-domain QA experiments, prior work \cite{dpr,rag,izacard2021fid} commonly discards the original evidence and evaluates on a standardized English Wikipedia corpus similar to NQ-Open. The ``filtered'' version ensures that all retained questions have at least one answer string present in the passage collection, yielding a cleaner evaluation set. Similar to \cite{realm,atlas}, we perform evaluation on the dev set.

\subsection{Methods Investigated}

\para{Baselines}
We compare LURE-RAG\footnote{Code available at \url{https://github.com/ManishChandra12/LURE-RAG}} against the following baselines:
\uls
\li \textbf{0-shot}: We use the base models as the baseline, i.e., without providing any retrieved context in the prompt.
\li \textbf{k-shot}:
We use the top-$k$ documents/passages retrieved by the first-stage retriever. This baseline helps to check whether the notion of relevance in IR translates to RAG directly.
\li \textbf{RePlug} \cite{replug}: The original RePlug framework optimizes a retriever by aligning retrieval scores with downstream LLM utility. However, RePlug has two practical limitations in our setting. First, it is incompatible with fixed or non-learnable retrievers such as BM25, since its learning relies on updating the retriever parameters. Second, it is resource-intensive, as the retriever optimization requires periodic index refreshes to incorporate updated document representations. To enable a fair comparison, we adapt RePlug into a black-box retriever setting: the retriever is fixed, and only a reranker is trained. Specifically, we retain RePlug’s KL-divergence–based objective, but apply it to align the reranker scores with downstream utility signals derived from the LLM outputs. This modified baseline preserves RePlug’s utility-driven supervision while removing its dependency on the retriever updates and index refreshes, making it directly comparable to our proposed LURE-RAG method. 
\ule

\para{Variants of LURE-RAG}
To evaluate the effect of ranking loss in isolation (i.e., without efficiency consideration), we employ the following variants of LURE-RAG. 
\uls
\li 
\textbf{LURE-RAG}: We employ Phi-3-mini-4K-instruct \cite{phi} as the LLM for obtaining utilities in order to train the LambdaMART reranker.
\li 
\textbf{UR-RAG}: We also introduce a dense variant, UR-RAG, where we replace LambdaMART with SBERT \cite{sbert}. SBERT has been widely adopted for semantic retrieval and reranking, and it has been extensively benchmarked \cite{beir,hofstatter2021efficiently} in retrieval settings. 
We fine-tune SBERT using the following loss function \cite{lambdarank,udr} to inject the ranking signal into the reranker.
\begin{equation}
    \pazocal{L}_{rank}(q) = \sum_{d_i, d_j \in D_q: U(q,d_i)>U(q,d_j)} \Big(\frac{1}{r(d_i)} - \frac{1}{r(d_j)}\Big) \log(1+e^{(s_{q,d_j}-s_{q,d_i})}),
\end{equation}

where $r(\cdot)$ is the ranking induced by $U(\cdot)$. If $U(q,d_i)>U(q,d_j)$, $d_i$ is ranked higher than $d_j$ and therefore, $r(d_i) < r(d_j)$.  When $d_i$ has a much higher rank than $d_j$, e.g, $r(d_i) = 1$ and $r(d_j) = 10$, the weight component of the loss will be high and the loss function will strongly draw $s(q, d_i)$ up from $s(q, d_j)$. This variant provides insights into the effectiveness-efficiency trade-off.
\ule

\subsection{Evaluation Metrics and Hyper-parameter Configurations} \label{subsec:metrics}

\textbf{Accuracy.} Datasets like NQ-Open and TriviaQA provide a range of potential answers for each query. Frequently, these answers are different variants of the same concept (e.g., ``President D. Roosevelt'' or ``President Roosevelt''). To evaluate the LLM accuracy on these datasets, which accept a range of valid answers for each query, we employ a binary assessment consistent with \cite{kandpal2023large,power_of_noise}. A response is deemed accurate if it contains at least one of the predefined correct answers; otherwise, it is inaccurate.

\noindent \textbf{F1.} F1-score is useful when answers involve partial overlap (e.g. multi-token answers), or when one wants to reward partial correctness or penalize missing parts of the answer. For a query $q$, precision and recall are defined as:
%
%
$P_q = |y_q \cap \hat{y}_q|/|\hat{y}_q|$ and 
$R_q = |y_q \cap \hat{y}_q|/|y_q|$ respectively. We then define the F1 score as:
$F1 = (1/|Q|)\sum_{q \in Q} \max_{y_q \in C_q}(2P_qR_q/(P_q+R_q))$,
where $C_q$ denotes the set of all available ground-truth answers for the query $q$.



LURE-RAG relies on the LLM predictions for the reranker training. To analyse the variations that may be caused due to this choice of LLM, we conduct our experiments on three different LLMs - each from different families. The objective is to analyse the variations in LURE-RAG's performance corresponding to different sizes of the models, and variations in characteristics of models across different families, thereby allowing to answer RQ-4. In particular, we use Meta's Llama-3.2-1B-Instruct \cite{llama32}, Microsoft's Phi-3-mini-3.8B-4K-instruct \cite{phi3} and Alibaba's Qwen2.5-14B-Instruct-1M \cite{qwen2.5}. Furthermore, we conduct two sets of experiments - one involving BM25~\cite{bm25} for sparse retrieval and the other involving Contriever~\cite{contriever} for dense retrieval.

In our experiments, we set the number of first-stage retrieved documents using BM25/Contriever ($N$) to 10 and RAG context size ($k$) to 5.
As a reranker for RePlug baseline and UR-RAG, we finetune SBERT all-MiniLM-L6-v2 \cite{sbert}. For fine-tuning SBERT, we use a training batch size of 32, a learning rate of 1e-5 and early stopping with a patience of 3 epochs. As an optimiser, we use Adam with weight decay. The optimal values of the hyperparameters are obtained via a grid search. For LDA, we use a random subset of 1M documents from the corpus and set the number of topics $M$ to 100 and the threshold for `top' topics (a) to 20.

\begin{table}[t]
\centering
\caption{A comparison between our proposed LURE-RAG approach and the baselines (using BM25 for retrieval) on the two datasets in terms of Accuracy (Acc) and F1-score. The best Acc and F1 values are bold-faced for each dataset and LLM combinations.}
\label{table:results_bm25}
\begin{adjustbox}{width=0.96\columnwidth}
\begin{tabular}{@{}l@{~~}l@{~~}l@{~~} rrrrrr@{}}
\toprule
& & & 
\multicolumn{2}{c}{\textbf{llama-1B}} & 
\multicolumn{2}{c}{\textbf{phi-3.8B}} &
\multicolumn{2}{c}{\textbf{Qwen-14B}} \\
\cmidrule(lr){4-5} \cmidrule(lr){6-7} \cmidrule(lr){8-9}
\textbf{Dataset} & \textbf{Rank-By} & \textbf{Method} & \textbf{Acc} & \textbf{F1} 
& \textbf{Acc} & \textbf{F1} & \textbf{Acc} & \textbf{F1} \\
\midrule
\multirow{6}{*}{NQ} 
  & \multirow{2}{*}{Relevance} & 0-shot & .1215 & .0616 & .1970 & .1026 & .2323 & .2518 \\
  &                                  & k-shot & .2876 & .1590 & .2676 & .1866 & .3548 & .3498  \\
\cmidrule(lr){2-9}
  & \multirow{4}{*}{Utility}   & RePlug & .2932 & .1894 & .2895 & .2007 & .3775 & .3700 \\
  &                                  & LURE-RAG       & .2891 & .1868 & .2822 & .1963 & .3712 & .3655 \\
& & LURE-RAG (w/o topics) & .2863 & .1829 & .2810 & .1961 & .3700 & .3627  \\    
  \cmidrule(r){3-9}
  &                                  & UR-RAG  & \textbf{.2999} & \textbf{.2068} & \textbf{.2933} & \textbf{.2088} & \textbf{.3822} & \textbf{.3854} \\  
\midrule
\multirow{6}{*}{TQA} 
  & \multirow{2}{*}{Relevance} & 0-shot & .2258 & .2137 & .2779 & .2214 & .3566 & .3587 \\
  &                                  & k-shot & .2841 & .2600 & .3486 & .3177 & .3976 & .3978 \\
\cmidrule(lr){2-9}
  & \multirow{4}{*}{Utility}   & RePlug & .3027 & .3017 & .3603 & .3321 & .4112 & .4125 \\
  &                                  & LURE-RAG       & .2968 & .3010 & .3569 & .3272 & .4057 & .4047 \\
&& LURE-RAG (w/o topics) & .2947 & .2973 & .3544 & .3261 & .4008 & .4001  \\
\cmidrule(r){3-9}
  &                                  & UR-RAG  & \textbf{.3111} & \textbf{.3104} & \textbf{.3676} & \textbf{.3472} & \textbf{.4186} & \textbf{.4239} \\
\bottomrule
\end{tabular}
\end{adjustbox}
\end{table}

\begin{table}[t]
\centering
\caption{A comparison between our proposed LURE-RAG approach and the baselines (using Contriever for retrieval) on the two datasets in terms of Accuracy (Acc) and F1-score. The best Acc and F1 values are bold-faced for each dataset and LLM combinations.}
\label{table:results_contriever}
\begin{adjustbox}{width=0.96\columnwidth}
\begin{tabular}{@{}l@{~~}l@{~~}l@{~~} rrrrrr@{}}
\toprule
& & & 
\multicolumn{2}{c}{\textbf{llama-1B}} & 
\multicolumn{2}{c}{\textbf{phi-3.8B}} &
\multicolumn{2}{c}{\textbf{Qwen-14B}} \\
\cmidrule(lr){4-5} \cmidrule(lr){6-7} \cmidrule(lr){8-9}
\textbf{Dataset} & \textbf{Rank-By} & \textbf{Method} & \textbf{Acc} & \textbf{F1} 
& \textbf{Acc} & \textbf{F1} & \textbf{Acc} & \textbf{F1} \\
\midrule
\multirow{6}{*}{NQ} 
  & \multirow{2}{*}{Relevance} & 0-shot & .1215 & .0616 & .1970 & .1026 & .2323 & .2518 \\
  &                            & k-shot & .2911 & .1782 & .2631 & .1835 & .3508 & .3401  \\
\cmidrule(lr){2-9}
  & \multirow{4}{*}{Utility}   & RePlug                & .2976 & .1853 & .2799 & .1994 & .3751 & .3696 \\
  &                            & LURE-RAG              & .2945 & .1811 & .2722 & .1949 & .3687 & .3669 \\
  &                            & LURE-RAG (w/o topics) & .2940 & .1798 & .2720 & .1904 & .3666 & .3664  \\    
\cmidrule(r){3-9}
&                              & UR-RAG                & \textbf{.3027} & \textbf{.1915} & \textbf{.2883} & \textbf{.2076} & \textbf{.3811} & \textbf{.3800} \\  
\midrule
\multirow{6}{*}{TQA} 
  & \multirow{2}{*}{Relevance} & 0-shot & .2258 & .2137 & .2779 & .2214 & .3566 & .3587 \\
  &                            & k-shot & .2834 & .2620 & .3509 & .3137 & .3966 & .3954 \\
\cmidrule(lr){2-9}
  & \multirow{4}{*}{Utility}   & RePlug                & .3111 & .3042 & .3582 & .3400 & .4010 & .4077 \\
  &                            & LURE-RAG              & .3059 & .2951 & .3635 & .3304 & .3994 & .4039 \\
  &                            & LURE-RAG (w/o topics) & .2994 & .2897 & .3628 & .3220 & .4010 & .4017  \\
\cmidrule(r){3-9}
&                              & UR-RAG                & \textbf{.3184} & \textbf{.3098} & \textbf{.3699} & \textbf{.3497} & \textbf{.4177} & \textbf{.4203} \\
\bottomrule
\end{tabular}
\end{adjustbox}
\end{table}

\section{Results} \label{sec:results}
Tables~\ref{table:results_bm25} and \ref{table:results_contriever} report the performance of our proposed method relative to the baselines on two open-domain QA benchmarks, Natural Questions (NQ) and TriviaQA (TQA), across three different LLMs of increasing model sizes --- llama-1B, phi-3.8B, and Qwen-14B.

\para{Comparison between relevance-based and utility-driven paradigms}
Across the dataset and LLM choices, we observe that utility-driven approaches consistently outperform the relevance-based ones. While k-shot prompting improves over 0-shot in the relevance-based setting (e.g., +0.16 Acc improvement for llama-1B on NQ using sparse retrieval), the gap between k-shot and utility-driven reranking remains substantial. This confirms that document ordering based solely on classical IR relevance is insufficient for optimal downstream performance, and utility-driven reranking better aligns retrieval with the LLM’s actual answer quality. Therefore, in answer to \textbf{RQ-1}, we conclude that utility-driven reranking leads to better downstream performance than traditional relevance-based retrieval across different LLMs and datasets.

\para{Effect of listwise ranking loss}
While RePlug remains a strong utility-driven baseline, we observe that UR-RAG significantly outperforms it across all settings. For example, on NQ with llama-1B using sparse retrieval, RePlug achieves 0.1894 F1, whereas UR-RAG achieves 0.2068, yielding noticeable improvements. Paired t-tests at 5\% significance level reveal that the improvements in the evaluation metrics obtained using UR-RAG are statistically significant across all dataset and LLM combinations (tests are conducted separately) compared to Replug. A key reason for this gain lies in the ranking loss employed by UR-RAG, which explicitly encourages the reranker to preserve correct document orderings according to the downstream utility. In contrast, RePlug does not incorporate any ranking-aware objective, relying solely on the KL divergence between the reranker scores and utility distributions. This difference in performance is observed because order of documents matter in RAG \cite{power_of_noise}. Therefore, in answer to \textbf{RQ-2}, we conclude that training a reranker with listwise ranking loss derived from LLM utility leads to better downstream performance.

\para{Comparison between dense and light-weight reranker}
LURE-RAG achieves results that are competitive with the dense reranker baseline (RePlug), while requiring only a lightweight LambdaMART model for training and inference. The F1-scores and accuracies for the NQ dataset using LURE-RAG (in sparse retrieval setup) are always above $97.8\%$ and $97.4\%$, respectively, of that achieved using RePlug. Similar numbers for TQA dataset are $98.1\%$ and $98.05\%$, respectively. These results indicate that a lightweight reranker, when guided by \textit{listwise ranking} utility-driven supervision provides a much more efficient alternative with only minor effectiveness trade-offs compared to a dense neural reranker not trained with a ranking loss. Therefore, in answer to \textbf{RQ-3}, we conclude that a lightweight LambdaMART-based reranker can achieve competitive performance compared to dense neural rerankers that are trained without ranking loss, while being more computationally efficient.

Along the expected lines, UR-RAG, which fine-tunes SBERT with a utility-driven ranking loss, yields the highest accuracy and F1 scores across all dataset-model combinations. For instance, on TQA with Qwen-14B (in sparse retrieval setup), UR-RAG achieves 0.4186 accuracy and 0.4239 F1, the strongest results in Table~\ref{table:results_bm25}. The F1-scores and accuracies for the NQ dataset using LURE-RAG are always above $90.3\%$ and $96.2\%$, respectively, of that achieved using UR-RAG. Similar numbers for TQA dataset are $94.2\%$ and $95.4\%$, respectively. These results indicate that when a lightweight reranker and a dense neural reranker, both trained using a ranking loss, are compared, the performance deterioration is larger.

Furthermore, we also see improvements across different LLMs. Note that we train the reranker using the utilities obtained from the phi-3 model, and reuse the same trained reranker for other LLMs as well. Therefore, in answer to \textbf{RQ-4}, we conclude that the reranker trained on one LLM's utilities can perform well when used with other downstream LLMs of different sizes.

\begin{figure}[t]
    \centering
    \begin{subfigure}[t]{0.49\textwidth}
        \centering
        \includegraphics[width=.99\linewidth]{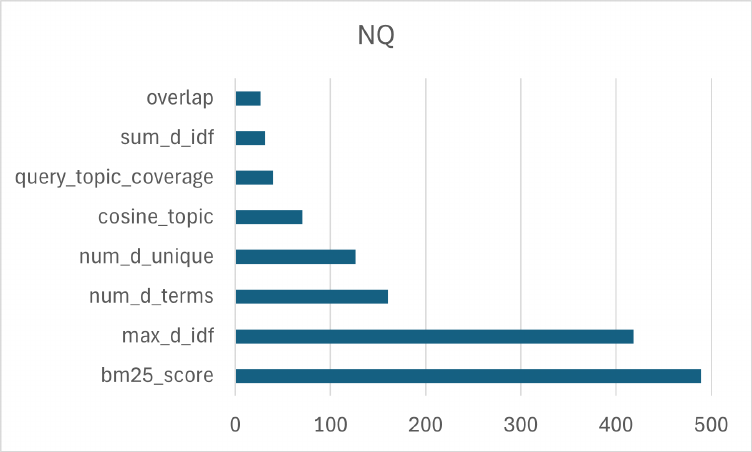}
        \caption{NQ}
    \end{subfigure}
    \begin{subfigure}[t]{0.49\textwidth}
        \centering
        \includegraphics[width=.99\linewidth]{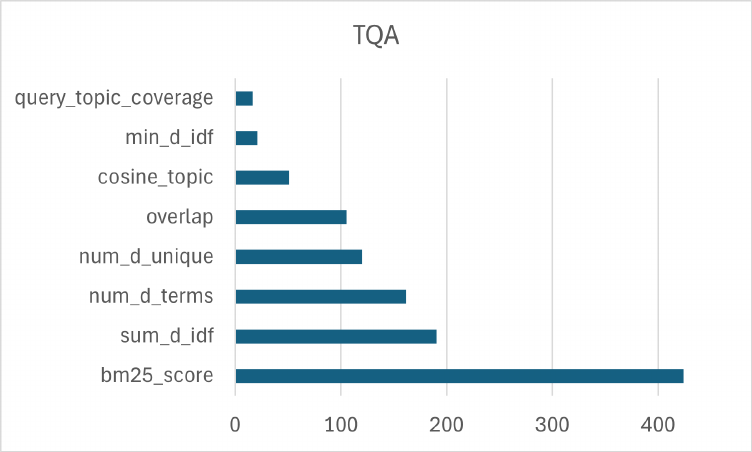}
        \caption{TQA}
    \end{subfigure}
    \caption{    
    Feature importance scores of the top-8 features obtained from the trained lambdaMART model for LURE-RAG.
    }
    \label{fig:feature_imp}
\end{figure}

\para{Interpretability}
The benefit of using a lightweight model like lambdaMART is that it's interpretable. To understand which signals drive the reranker’s decisions, we analyse the feature importance scores from the trained LambdaMART models (see Figure~\ref{fig:feature_imp}). Feature importances are computed using the gain metric, which measures the relative contribution of each feature to the model’s decision splits. For every split in the trees, the ``gain'' (improvement in the objective function) is calculated. The total gain for a feature is summed across all splits where it is used, providing a measure of its importance. The most influential feature is the BM25 similarity score (bm25\_score), which dominates the learned model. This suggests that classical lexical similarity remains a strong signal for utility-driven reranking. Beyond this, the number of (unique) document terms and document-level IDF statistics also exhibit notable importance, indicating that documents containing rare, high-value terms often align better with downstream LLM utility. Moreover, query-document term overlap and features derived from LDA provide additional, though relatively weaker, signals.

\para{Ablation study}
To examine the contribution of the LDA topic features, we conduct an ablation study by removing them from the feature set. As expected from the feature importance analysis (Figure~\ref{fig:feature_imp}), topic features exhibit relatively low importance compared to lexical and BM25-based features. Consistently, we observe only marginal degradation in performance across LLMs when these features are excluded, as shown in Tables~\ref{table:results_bm25} and \ref{table:results_contriever} corresponding to `LURE-RAG (w/o topics)' rows. This result suggests that while topic-level information can provide additional semantic signal, the reranker primarily relies on stronger lexical and retrieval-based signals. Nonetheless, the inclusion of LDA features does offer small but consistent improvements.

\section{Conclusions}
In this work, we proposed LURE-RAG, a lightweight utility-driven reranking framework that improves retrieval-augmented generation by incorporating a ranking-aware objective, offering a more efficient alternative to existing approaches. We further introduced UR-RAG, a dense variant that achieves the strongest results, significantly improving accuracy and F1 by up to 3\%. Our results demonstrate that utility-driven reranking using principled ranking losses provides a powerful and practical approach to aligning retrieval with downstream generation tasks in retrieval-augmented generation. In future work, we aim to explore richer feature representations including semantic signals.

\begin{credits}
\subsubsection{\discintname}
The authors have no competing interests to declare that are relevant to the content of this article.
\end{credits}



%
%
%
\bibliographystyle{splncs04}
\bibliography{custom}
%




\end{document}